# The road to safety- Examining the nexus between road infrastructure and crime in rural India

Ritika Jain (Corresponding author)

Assistant Professor

Centre for Development Studies, Trivandrum, India

E-mail: ritika@cds.edu

Shreya Biswas

Assistant Professor, Department of Economics and Finance

Birla Institute of Technology and Science, Pilani, Hyderabad Campus, India

E-mail: shreya@hyderabad.bits-pilani.ac.in

**Abstract**

*This study examines the relationship between road infrastructure and crime rate in rural India using a nationally representative survey. On the one hand, building roads in villages may increase connectivity, boost employment, and lead to better living standards, reducing criminal activities. On the other hand, if the benefits of roads are non-uniformly distributed among villagers, it may lead to higher inequality and possibly higher crime. We empirically test the relationship using the two waves of the Indian Human Development Survey. We use an instrumental variable estimation strategy and observe that building roads in rural parts of India has reduced crime. The findings are robust to relaxing the strict instrument exogeneity condition and using alternate measures. On exploring the pathways, we find that improved street lighting, better public bus services and higher employment are a few of the direct potential channels through which road infrastructure impedes crime. We also find a negative association between villages with roads and various types of inequality measures confirming the broad economic benefits of roads. Our study also highlights that the negative impact of roads on crime is more pronounced in states with weaker institutions and higher income inequality.*

**Declarations**

**Funding:** Authors declare that did not receive any funding for carrying out this study.

**Conflicts of interest/Competing interests:** The authors declare that they have no conflict of interest.

**Availability of data and material:** Data used for the study is available freely on

http://www.icpsr.umich.edu

# The road to safety- Examining the nexus between road infrastructure and crime in rural India

**Abstract**

*This study examines the relationship between road infrastructure and crime rate in rural India using a nationally representative survey. On the one hand, building roads in villages may increase connectivity, boost employment, and lead to better living standards, reducing criminal activities. On the other hand, if the benefits of roads are non-uniformly distributed among villagers, it may lead to higher inequality and possibly higher crime. We empirically test the relationship using the two waves of the Indian Human Development Survey. We use an instrumental variable estimation strategy and observe that building roads in rural parts of India has reduced crime. The findings are robust to relaxing the strict instrument exogeneity condition and using alternate measures. On exploring the pathways, we find that improved street lighting, better public bus services and higher employment are a few of the direct potential channels through which road infrastructure impedes crime. We also find a negative association between villages with roads and various types of inequality measures confirming the broad economic benefits of roads. Our study also highlights that the negative impact of roads on crime is more pronounced in states with weaker institutions and higher income inequality.*

JEL codes-R42, O18, K41, R10

# 1. Introduction

Assurance of safety is one of the fundamental aspects of civil society (Iyer et al., 2012). Living with a feeling of unsafety and the fear of being subjected to crime leads to perverse impacts on the quality of life (Dutta and Hussain, 2009). People tend to become more risk-averse when facing significant changes in the external environment, such as a natural calamity, civil unrest, pandemic or violent crime (Brown et al., 2019). Moreover, a crime-induced rise in risk aversion has adverse implications by restricting human mobility and reducing access to job and educational opportunities (Cook et al., 2013; Dutta and Hussain, 2009). Accordingly, even the Sustainable Development Goal (SDG) 16 highlights that reducing various levels of organized crime is an essential tool to achieve other SDGs[1]. The extant literature underscores several other adverse economic effects of crime. Crime activities increase uncertainty, are related to inefficient resource allocation, and deters investment (Detotto and Otranto, 2010). There is a negative correlation between crime, economic growth, and employment (Detotto and Pulina, 2013; Goulas and Zervoyianni, 2013).

Although important for developed economies, the adverse impact of crime is of utmost importance in developing economies already plagued with low growth, low investment trap, and uncertain economic environment. Further, due to informal markets, weak institutions and poor quality of infrastructure, the probability of being caught and convicted of a crime are lower in developing countries than the developed ones. Studies also document that countries that are characterised by higher inequality have a higher incidence of crime due to the following reasons (Ehrlich, 1974; Fajnzylber et al., 2002). Firstly, in societies with a high degree of inequality, the legal wage for low-skilled workers may be too low compared to expected earnings from indulging in illegal activities. Second, the cost of crime is the combination of the probability of being caught and prison time. Again when inequality is high, one may argue that the quality of life within and outside the prison may not differ substantially, making crime a high expected return and low-risk activity compared to low inequality scenarios. Since developing countries have higher inequality (Van der Hoeven, 2019), the vulnerability of individuals and exposure to crime is likely to be high in this setup. Frequent criminal activities may affect productivity in developing economies hindering life. Thus, research on factors and interventions that may control and deter crime may be highly relevant for the developing world.

---

[1] Source- https://www.unodc.org/unodc/en/sustainable-development-goals/sdg16_-peace-and-justice.html, https://www.swp-berlin.org/publications/products/comments/2015C45_vrr_bsh.pdf (Accessed on July 11, 2021)

Becker (1986) pioneered the economics of crime and suggested that criminals were rational agents deciding whether to indulge in criminal practices based on their benefits and costs.

Following Becker (1986), a large body of literature has emerged that has explored the determinants of crime (Fajnzylber et al., 1998; Cahill and Mulligan, 2003; Imrohoroglu et al., 2006; Bunanno and Montolio, 2008). These papers identify several attributes such as unemployment rate, urbanization extent, the fraction of foreigners, previous incidence of crime and quality of institutions. On the other hand, sociological literature focuses on how the social theory of relative deprivation may be one of the significant determinants of crime (Merton, 1968; Blau and Blau, 1982; Bernburg et al., 2009, O'Mahony, 2018). The theory posits that more impoverished and more unequal societies have higher crime counts due to people feeling deprived relative to their peers. Besides these socio-economic factors, spending on road infrastructure may also influence crime rates (Hughes, 1998). Our paper also attempts to examine how building road infrastructure may impact criminal activities.

We explore various channels through which roads may influence crime. First, local development through roads may lead to better employment opportunities. A revisit to Becker's (1968) model then implies that the opportunity cost of crime rises with better employment opportunities. As a result, individuals may substitute their time spent on crime with formal employment. Hence, Becker's (1968) framework implies that building road infrastructure should impede and deter crime. However, if the economic benefits of employment due to roads disproportionately favour the skilled and endowed individuals more, the unskilled ones may still indulge in criminal activities. In some instances where the benefiting group forms a minuscule share of the population, it may lead to a rise in criminal activity.

Another channel that determines how road infrastructure may influence crime stems from the infrastructure development implementation itself. Roads reduce the time costs and increase mobility, both critical for criminal activities and economic activities. A well-connected road network may catalyze movements of criminals to potential hot spots with ease. These contrasting channels provide an interesting backdrop to test the empirical validation of how building road infrastructure may affect crime in developing economies.

Against this background, we attempt to examine how building road infrastructure impacts crime in rural India. The focus on rural India stems from the weak infrastructure and scarce non-farm opportunities (Jha, 2006)[2]. With heavy reliance on agriculture for employment, infrastructure development in rural parts of India has been slow. Despite receiving attention in several development plans and policies since independence, the slow pace had been persistent until the late nineties. In 2000, the central government of India introduced the Pradhan Mantri Gram Sadak Yojna (PMGSY) that aimed to connect all villages with an all-weather pucca road in a phased manner. Rule-based population cutoffs determined the sequence of phases. However, in 2011, the PMGSY rollout was extended to all the villages in India.

We use data from the India Human Development Survey (IHDS) conducted in two waves- 2004-05 and 2011-12. Among several socio-economic attributes, the survey explored whether the household faced any type of criminal activity in the last twelve months. We use multiple measures of crime as our dependent variables. Both waves of IHDS also have a separate questionnaire for village-level amenities, population composition and occupation structure, among other attributes. Using information from the village questionnaire on whether the village was accessible through an all-weather pucca road or kaccha road or was inaccessible, we construct our focal variable- the presence of a pucca road in the village.

We employ an instrumental variable estimation strategy to account for omitted unobservable factors that may simultaneously influence road and crime measures. We find that households living in villages connected with an all-weather pucca road experience 5% less criminal activities than households living in villages without it. Our effect size doubles when we control population composition, inequality and income uncertainty at the village level. A closer examination of the specific type of criminal activity reveals that the effect is limited to types of crime that have a higher possibility of happening outside home premises- female harassment and burglary.

We explore several channels that may drive our main findings. We posit that roads as a deterrent to crime may work through two channels- direct effect of better street lighting, higher likelihood of bus service and increased employment opportunities. Additionally, it will also bear the indirect benefit of higher income for the households and lower inequality at the village level. We test these channels by examining the impact of a pucca road on street lighting, and bus stops in the village, employment and income status of households. We find strong evidence that households in villages with better-connected roads have greater access to public programs related to street lighting and bus services. Additionally, we also find evidence for increased employment, higher income and equal land distribution for villages with better roads. These results outline the primary channel through which roads reduce crime.

---

[2] From the perspective of crime, crime is generally viewed as a byproduct of poverty, inequality and urbanization. According to the Indian Human Development Survey, crime rates (defined as the share of households that were subjected to some type of crime) in urban India was 7% between 2004-05 to 2011-12. During the same period, rural India witnessed a crime rate from 6.1%. Thus, crime in India is not substantially different between the sectors.

We extend our model in two broad ways. These extensions are based on institutional factors and pre-existing socio-economic conditions in the state where the village is located. In the first extension, we attempt to examine if our impact is uniform for states with better quality of institutions vis-à-vis states that do not. We use measures that capture the efficacy of crime deterrence and management at the state level and divide our states according to high and low categories. We find that road infrastructure reduces crime only in states that have a lower quality of institutions. This outlines the critical importance of roads in helping these states catch up with better institutions. As a second extension, we test if our effects are conditioned by the level of inequality and the coverage of public employment programs at the state level. Again, we find that our impact is limited to relatively more unequal states and that have better coverage of public employment program. These results underline the importance of building road infrastructure in rural India.

The paper is organized as follows: Section 2 lays down the relationship between crime and road infrastructure. We discuss the Indian experience in Section 3 and data and descriptive statistics are presented in Section 4. The econometric methodology is discussed in section 5. We present our results in Section 6 and conclude in section 7.

## 2. Relationship between crime and road infrastructure

**2.1 Economic effects of road infrastructure**

The need for infrastructure is of paramount importance in developing countries due to weak institutional factors (Sawada et al., 2014). These countries invest a large amount of resources in restructuring and building a broad and well-connected road network. An established road infrastructure setup leads to a wide range of impacts on urbanisation, population and environment. These impacts may be beneficial or harmful depending on the broader context (Khanani et al., 2021). For instance, the building of roads and transportation services accompany peripheral residential development, the emergence of commercial establishments and other forms of spatial segregation. Additionally, infrastructural development also enhances mobility, consequently reducing barriers to labour force participation (Akee, 2006; Lei et al., 2019), improving access to schools (Adukia et al., 2020) and health care facilities (Aggarwal, 2021).

The socio-economic benefits of roads are driven by the primary channel of reduced transportation costs and higher mobility. The resultant ease of access may open up job opportunities and public services that were previously inaccessible. For instance, Khandker (1989) finds that government investment on roads in Indian districts between 1961 and 1981 was associated with new non-farm employment and higher wages. Several recent studies have also documented the Indian experience of the creation of non-farm jobs due to rural road infrastructure investment (Aggarwal, 2018; Asher and Novosad, 2016).

However, roads may have some adverse outcomes as well. Past studies have examined the direct negative effect of environmental degradation, higher chances of landslides and road accidents (Forman and Alexandar, 1998; Paul and Meyer, 2001; Slabbekoorn and Peet, 2003; Coffin, 2007). Besides these, it may also have non-uniform impacts across various sections of society. For instance, people who own land and vehicles may use the roads to their advantage, whereas the landless may not benefit as much. Consequently, road infrastructure development may lead to rising inequality with the impact of roads favouring the rich and endowed more than the poor (Aggarwal, 2018). However, Ferriera (1995) argues that if infrastructure investment in underdeveloped areas increases connection between core economic activities, it may lead to more productive opportunities for the poor that may reduce inequality. In summary, the direction of the impact of roads on inequality is ambiguous depending on the context and the country.

**2.2 Roads and crime**

An additional aspect related to road networks that has received relatively less attention is criminal activity. Criminal activity is as spatially segregated as economic opportunities. The ease of mobility and better connectivity due to road infrastructure that benefits economic opportunities is also relevant for illegal activities. Further, the choice of committing crime stems from limited viable economic opportunities (Becker, 1968). In a developing country like India, where 11.90% of the population is unemployed[3], and a mere 20% of employed individuals are employed as waged and salaried workers[4], most individuals face a lack of dignified job opportunities. While there are several contributing factors to the employment situation in India, weak and inadequate infrastructure exacerbates it. Consequently, it can be related to a higher possibility of low or unskilled individuals indulging in criminal activities in given lack of job opportunities along with low wage returns. Against such backdrop, investing in road infrastructure may be an effective tool in reducing barriers to entering the labor force and reducing criminal activity.

---

[3] This is the unemployment rate in India as of May, 2021. Source- https://unemploymentinindia.cmie.com/
[4] This is the data for May 2019. Source- https://timesofindia.indiatimes.com/business/india-business/why-bad-employment-is-a-bigger-problem-than-unemployment/articleshow/70099901.cms

Further, if infrastructure benefits the elite and non-poor more than the poor and economically vulnerable sections of the society, it may make the society more unequal. Road infrastructure may then increase crime as a consequence of a rise in inequality. Becker (1968) explains that to catch up with high-income individuals, crime is an easier tool for low-income individuals than the low returns from the labor market. Several papers have extended this model (Ehrlich, 1973; Block and Heineke, 1975; Chiu and Madden, 1998). Bourguignon (2000) also finds a positive association between observed inequality and levels of crime. Bourguignon (2001) revisits this issue with a particular focus on urbanization and reiterates that crime is a byproduct of uneven economic development or processes. Inequality and poverty, even if transitory, have large and persistent societal losses through crime.

A similar positive association between crime and inequality is presented in Merton (1938). Discussed in the sociological literature, the strain and disorganization theory by Merton (1938) posits that individual alienation due to low income, marginalized status, or discrimination may lead to indulging in criminal activities. The empirical evidence of these theories, however, remains inconclusive. While Blau and Blau (1982) and Bourguignon et al. (2003) find support for it, Land et al. (1990), Kelly (2000) and Kang (2016) find an insignificant relationship between crime and inequality. Bourguignon et al. (2003) investigate the relation for the seven largest cities in Colombia and find that probability of being a criminal was higher for individuals living in households that had a per capita income below 80 percent of the mean. In contrast, Kang (2016) emphasizes that crime is primarily driven by economic segregation instead of within neighborhood inequality.

A second dimension to the relationship between road infrastructure and criminal activities is related to the physical network of roads. According to criminology literature, road infrastructure may work as a skeletal structure to criminals that may aid in identifying and easily accessing the hot spots (Davies and Johnson, 2015). This may perpetuate criminal activities in areas that are well connected. However, one may argue that infrastructure activities require casual employment that benefits the unskilled category of the local population. Hence, considerable investments in infrastructure projects like road construction may boost employment among the poor and unskilled, leading to lower criminal activities.

The above discussion highlights the possibility of the association between road infrastructure and criminal activities to move in either direction. Against these contrasting channels, we attempt to evaluate the Indian experience of the impact of road infrastructure on crime.

### 3. The Indian experience

**3.1 Crime procedures and the issue of underreporting**

There are two possible ways to gather information on criminal activities in India- crime records maintained by the respective nodal agency in the country and information collected from victims directly through a survey. The former source of information is layered by several agents between the crime scene and the nodal agency. The Code of Criminal Procedures deals with criminal cases and procedures in India. According to this code, crime-related information reported to the police needs to be compiled and written in a First Information Report (FIR), then signed by the victim. Based on the FIR, the police start investigating the crime and making arrests if necessary. Once the police forces have arrested an accused individual, they are produced before the magistrate charges the person with the specified crime or releases them. Most crimes such as extortion, theft, harassment, murders, terrorism and so on are punishable under the Indian Penal Code.

National Crime Records Bureau (NCRB) is the nodal agency in India that collects, compiles and disseminates information related to crime. Published by the Ministry of Home Affairs, NCRB records annual data on various aspects of crime at the state and district levels of India. The primary source of information of this data is the FIR that is filed with the police force. The information is collated at the District Crime Records Bureau and sent to the respective State Crime Records Bureau from the police station where the FIR is filed. Finally, the NCRB consolidates all this information and compiles it.

Using police records of crimes from NCRB may have several limitations in capturing the pattern of actual crime, since various crimes in India go unreported due to poor quality of infrastructure, weak institutional factors and social stigma. In fact, reporting of crime is a problem across countries- according to the International Crime Victim Survey data only 40 percent of committed crimes are reported at the global level. However, under-reporting of crime is more pervasive in developing economies. Ansari et al. (2015) report that people in India do not report crimes due to the paucity of police stations, lack of awareness and inadequate trust in the criminal justice system. The study also documents that while the trend of violent crimes such as murders in India is not very different from the developed world, other criminal activities such as burglary and theft have a higher likelihood of not being reported as compared to the rest of the world[5]. Further, under reporting of crime in India, this issue is a dual consequence of victims choosing not to report and police deciding not to record it. While police records are valuable information, using victim-reported crime may reduce the scale of an understatement. Such data is gathered from surveys directly by asking respondents if they faced any crime in the past year. Prasad (2013) documents that while police-recorded crime patterns are well represented in Indian regions where institutions function efficiently, the difference in the two types of crime measures is enormous in the rest of the country.

*Road infrastructure in India*

Despite experiencing high economic growth in the early 2000s, India has been grappling with a weak and inadequate infrastructure network. While necessary for urban areas, access to better infrastructure is critical for poverty alleviation and economic development in the rural sector. Within the infrastructure sector, roads have been at the forefront of economic development in India, with rural road development plans receiving attention since independence. Despite receiving considerable attention, a lack of planning, improper design and low monitoring lead to several deficiencies in the rural road network (Samanta, 2015). Inadequate embankment and poor drainage network implied that most of these roads were not accessible during rough weather.

Against this background, a centrally sponsored scheme, PMGSY, was launched in 2000. The scheme's primary objective was to provide rural habitations (defined as a cluster of the population that resides at the same location along the lines of a hamlet) with an all-weather pucca road within 500 meters. While it was a scheme introduced by the central government of India, state and local governments were active participants in the implementation of the project. PMGSY was implemented according to population criteria in a phased manner. Villages that had a population of 1000 or more were prioritized in the first phase with few exceptions. The second phase involved villages with a population of 500 and finally the third one for villages with 250[6]. However, in 2010 the scheme was universally opened to all villages. Additionally, the roads being built were to be connected with the core network of roads within the state.

---

[5] These are deduced from two rounds in which select Indian cities were included in the International Crime Victim Survey in 1992 and 2003. Ansari et al. (2015) discusses this issue in detail.

In 1951, a mere 20% of Indian villages had access to an all-weather road[7] that increased to 60% in 2000 (Lei et al., 2019). As of 2019, the access has spread to 73% of Indian villages[8]. This broad coverage of road networks reflects that road infrastructure since the 2000s has grown. Using measures based on the PMGSY data to capture road infrastructure quality provides an accurate picture of capturing how uniform the growth has been across various regions.

## 4. Data and variables

We use the India Human Development Survey (IHDS), a nationally representative survey of more than 40,000 households. With comprehensive coverage of socio-economic variables such as health, education, gender relations, social networks, crime, confidence in institutions and so on, the IHDS dataset is well suited for addressing the impact of road infrastructure and crime. The survey was conducted twice, wherein the first wave was in 2004-05 and the second wave was in 2011-12. Thus, the dataset consists of a household-level panel for two years.

IHDS also collected information at the village level in both the waves from focus group discussions among village officials, people in business, and similar people in the village. The information spans various issues such as infrastructure, public programs, occupation structure, and population composition. We merge this village-level information with the household questionnaire. Since the village level information is collected only for rural areas, we drop all households dwelling in the urban sector[9]. Our final data set comprises more than 27,000 rural households for 2004-05 and 2011-12.

The dependent variable for our analysis is criminal activities experienced by victims (households) in the recent past. IHDS has four questions on whether households faced burglary, threats, female harassment and breaking into homes or not. Based on these criminal activities, we construct two measures of crime- (i) a simple average of each of the four types of crime and (ii) a dummy variable that takes a unit value if anyone in the household is a victim of either of the four types of crime and zero otherwise. To test the robustness of our model, we also use each of the types of crime as dependent variables in separate models.

---

[6] There were few exceptions to the rule. For instance, if a habitation with less than 1000 population lies on the straight path of a road that was built for a habitation with higher than 1000.
[7] Source- https://niti.gov.in/planningcommission.gov.in/docs/aboutus/committee/wrkgrp12/transport/wgrep_rural.pdf
[8] Source- https://theprint.in/india/governance/piyush-goyal-hails-indias-newly-connected-villages-but-26-still-await-pucca-roads/186739/
[9] An additional reason for focusing on rural areas is that PMGSY scheme was introduced only in rural India. We discuss this in detail in the next few paragraphs.

Our interest variable is related to road infrastructure in Indian villages. An ideal approach would have been mapping the PMGSY scheme implementation data with our dataset. However, due to the lack of village identifiers in IHDS data, we are unable to use the implementation of the PMGSY scheme at the village level. As an alternate resort, we exploit the stock nature of road infrastructure and use measures from the IHDS directly to measure the connectivity of villages with all-weather pucca roads. The village questionnaire asks whether the village is connected with a pucca (all-weather) road, a kaccha road or is not connected by a road at all. We use this information to construct a dummy variable for villages that are connected by an all-weather road. Consequently, villages not connected by any road or connected by a kaccha road get a zero value.

We use a set of control variables at the village and household levels that may influence crime. These variables broadly span across the presence of police stations in the village, confidence that households have in institutions critical to crime deterrence and management, land distribution in the village between the largest and the rest of the caste and religion groups and other socio-economic variables. We present the definition, measures and basic summary statistics of each of these variables in Table 1.

*<Table 1>*

As a first step in measuring aggregate patterns between road infrastructure and crime, we present how various economic outcomes differ between villages with pucca roads and villages without them. We compile the results in Table 2.

*<Table 2>*

Table 2 presents that most crime measures have a lower value for households dwelling in villages with an all-weather pucca road than households that reside in villages without it. Further, we find that villages with pucca roads also have a higher probability of getting street lights through a public program. Similarly, this group of villages also have a bus stop that is closer to the village than the group of villages that do have a pucca road. Finally, families dwelling in villages with well-connected roads exhibit better labour force participation rates and higher family income. These patterns lean towards the possibility of rural road infrastructure being effective in tackling crime in India. We explore this further in the next section, wherein we discuss our identification strategy.

## 5. Identification strategy

We aim to examine the impact of road infrastructure on crime. To meet this objective, we estimate the following model-

$$Crime_{hvdt} = \alpha_1 + \beta_1 Road_{vdt} + X_{hvdt}\gamma_1 + \varepsilon_{1hvdt} \text{------------------------------ (1)}$$

where $h$, $v$, $d$ and $t$ denote household, village, district and time respectively. $Crime_{hvdt}$ is the measure of crimes that household $h$ faces in the $v^{th}$ village, $d^{th}$ district and $t^{th}$ year. $Road_{vdt}$ is a dummy variable that indicates if the $v^{th}$ village is connected with an all-weather pucca road or not. $X_{hvdt}$ is a vector of household confounders, including religion, caste, and other economic variables. We also control for the district, state, and year to account for administrative quality, job opportunities, state specific police expenditure, and over time change in crime trends. $\varepsilon_{1hvdt}$ is the error term that captures the impact of all unobserved omitted factors.

For obtaining estimates of the impact of road infrastructure on crime, road infrastructure must be exogenous in our estimation model. However, the possibility of a set of unknown variables influencing both crime and road infrastructure may lead to an omitted variable bias and consequently render our road infrastructure variables endogenous to criminal activity. This requires using an instrumental variable estimation strategy wherein the first stage we estimate road infrastructure measures using a set of exogenous variables used in (1) and an additional instrumental variable. We estimate the following equation as the first stage of our model-

$$Road_{vdt} = \alpha_2 + \beta_2 PG_{vdt} + X_{hvdt}\gamma_2 + \varepsilon_{2hvdt} \text{-------------------------- (2)}$$

where all notations denote the same variables as in equation (1). The instrumental variable, denoted by $PG_{vdt}$, is based on the rationale that it affects road infrastructure but has no direct impact on criminal activities faced by households in that village. Since provisioning of different types of public goods, is correlated with each other at the village level, we use the proportion of households with access to piped drinking water as an instrument for building all-weather roads in villages (Banerjee and Somanathan, 2007). We posit that piped drinking water access and providing road infrastructure may have a strong positive association, but access to piped drinking water does not have an apparent effect on crime incidence. The predicted values from model (2) is then used as an explanatory variable in the second stage, denoted by eq (1). In the presence of valid and relevant instruments, this estimation yields causal impact of road infrastructure on crime. We discuss our findings in the next section.

## 6. Results

### 6.1 Main findings

Table 3 presents the estimates of the impact of road infrastructure on criminal activity. In Models 1, 2 and 3 we present the OLS estimates. The second stage of the instrumental variable regression models are presented in Models 4, 5 and 6.

<Table 3>

Table 3 demonstrates that the presence of pucca roads is negatively associated with crime. Since the 2SLS results control endogeneity, we use Models 4, 5 and 6 to discuss our main findings. According to Model 4, villages connected with pucca roads have a 4.7 percent lower incidence of criminal activities than villages not connected with pucca roads.

As a next step, we try to consider the focal employment generation activity in rural India- agriculture and the related income uncertainty. According to the Periodic Labor Force Survey of 2018-19, 58% of rural employment in India is generated from agriculture[10]. Hence, certainty in income for rural households will be closely aligned with prices and wages in the agricultural market. To account for that, we include the seasonal wage difference between the agricultural harvest and non-harvest time of the year in Model 5. Further, to capture unequal land distribution within the village, we also include the difference in land ownership between the largest and the rest of the religious and caste categories. Finally, we account for the presence of migrants within the village using a dummy variable. Model 5 depicts that our effect size increases to 12% as compared to Model 4.

In Model 6, we broaden our definition of crime index to a dummy variable that takes a unit value if households have faced any type of criminal activity. We also include all the village level exogenous covariates to control for unequal distribution of land and income uncertainty. Compared to Model 4, we find that the coefficient magnitude of pucca road increases when we use this measure of crime. Broadly, we find that road infrastructure in India leads to a negative impact on crime.

---

[10] Source- https://www.thehindubusinessline.com/opinion/why-agriculture-sectors-share-in-employment-is-declining-in-rural-india/article32900228.ece

We find that criminal activity is associated with households' confidence in various institutions in the country. Specifically, crime is negatively associated with confidence in public institutions such as police, panchayat and courts. Additionally, the presence of a police station in the village is also associated with fewer criminal activities. Finally, we control for household-level attributes that may affect being affected by some criminal activity. To account for unobserved effects, we include fixed effects at the district, caste, religion and year level. The impact of pucca roads on crime in Models 4, 5 and 6 of Table 3 requires that the instruments are valid and exogenous. We present the estimates of our instrument, proportion of households in a village that has access to piping water in Table 4. In the bottom panel of Table 4, we present the tests of instrument validity for it.

<Table 4>

Table 4 indicates that households with piped drinking water are positively associated with households that have an all-weather pucca road confirming that patterns of provisioning of different types of public goods are correlated with each other at the village level (Banerjee and Somanathan, 2007). Further, the bottom panel confirms that our instrument is valid and relevant.

**6.2 Robustness test**

**6.2.1 Effect of road infrastructure on various types of criminal activities**

It is imperative to examine the effect of pucca roads on different types of criminal activities- theft, attack or threats, female harassment and breaking into homes. Of the four types of activities, breaking into homes is the only type of crime that one is subjected to at the residential facility. We compile these results in Table 5.

<Table 5>

Interestingly, we observe that pucca roads have a negative and significant impact on the propensity of being robbed and for females to be harassed (Table 5). However, it has a positive effect on burglars breaking into homes and is insignificant for getting attacked. These overall results confirm that robbing and harassment have a higher incidence on poorly built paths. This result aligns with Mahajan and Sekhri (2020) wherein building in-home toilets reduces the risk of violent crimes against women. Along similar lines, building an all-weather road makes it easier for women to access different places than walking in secluded areas with no roads. A weak but qualitatively similar argument may be plausible for being robbed in rural areas with no well-connected roads. In contrast, the rise in burglars breaking into homes due to pucca roads aligns well with Davies and Johnson (2015). This affirms that road infrastructure may work as a skeletal structure to criminals that may aid in identifying and easily accessing the hot spots.

### 6.2.3 Plausibly exogenous model

Our instrument validity tests confirm that our Kleinbergen-Papp F statistic is above the thumb rule of 10. While our instruments are relevant, the assumption of exogeneity is challenging to assess. To account for this aspect, we use the Conley et al. (2012) 'plausible exogenous' estimation of our main effects. It enables us to examine the impact of road infrastructure on crime even if the instrument is not entirely exogenous. First, we estimate the reduced form equation of crime on the instrument and other controls but excluding the endogenous variable and obtain the estimate for the lower bound of gamma[11] (Azar et al., 2021; Das, 2018; Biswas and Das, 2021). Using gamma values between 95% of estimated gamma min and zero, we examine if the effect of road infrastructure is still negative. We compile the results in Table 6.

*<Table 6>*

Table 6 provides evidence that road infrastructure is negatively associated with crime and this finding is robust to violation of the strict exogeneity assumption of the instrument. Tables 3-6 affirm that road infrastructure is an effective tool in crime deterrence in rural India.

### 6.3 Potential channels driving the main effect

We explore the potential pathways through which road infrastructure reduces crime. We postulate that there may be direct and indirect benefits of road infrastructure in the village that may reduce crime. Perkins et al. (2015) find that reducing streetlights in England leads to an increase in burglary in more deprived areas. Similarly, Desai and Vanneman (2019) find that road infrastructure is followed by public bus services that may enable easier access to farmers to get their produce to nearby markets, the older children to take admission in schools and colleges further away get to health centers and hospitals easily. With easier accessibility, villagers can commute greater distance to find better-suited job opportunities (similar to the findings of Khandker, 1989; Aggarwal, 2018; Asher and Novosad, 2016).

---

[11] Gamma is the association between the instrument and the endogenous variable and the reduced form regression estimate acts as lower bound for gamma.

We use street lighting, bus stops, and the share of adults in the age group of 18-60 years from a household who are employed as the three indicators that measure the direct impact of road infrastructure. We use this information and examine if an all-weather pucca road is positively associated with these outcome variables. We compile this information in Table 7.

*<Table 7>*

Column 1, 2 and 3 of Table 7 confirms that building an all-road pucca road in a village has a positive and significant impact on villages getting streetlights as part of a public programme, higher likelihood of having a bus service in the village itself and a higher share of working adults within the household. These findings corroborate the direct effects of building road infrastructure in rural India.

These economic benefits have larger indirect effects that favor a higher quality of life which consequently reduces criminal activities. Hence, we examine if building pucca roads impacts income on resident households (Column 4 of Table 8) and find that households dwelling in villages with pucca roads have higher income. Since inequality and relative deprivation are central to crime's social disorganization theory, we also examine the impact of pucca roads on two measures based on land ownership. We use the squared difference of share of land owned by the general and the marginalized caste category (Column 5)[12]. We construct a similar measure for religion (Column 6). Results from columns 5 and 6 confirm that pucca roads have a negative impact on these measures.

These results confirm the broader findings of Aggarwal (2018), Asher and Novosad (2016) and Adukia et al. (2020). Better road infrastructure leads to easier connectivity and access to economic opportunities that were previously unavailable to them. This is reflected as a rise in the employment of adults in the working-age population of households. Finally, households become better off as depicted with a rise in income levels. Unlike Aggarwal (2018) we do not find that roads exacerbate inequality. In fact, our findings indicate that roads are related to a decline in inequality in villages as reflected through a reduction in inequality in landholding, one of the most valuable assets in the rural context. These channels broadly confirm that reduction in criminal activity due to road infrastructure in rural areas is due to the direct effect of better street lighting and the indirect effect of reduced barriers to better opportunities.

---

[12] This measure is based on the concept of variance that gives higher weight to observations further away from mean.

## 6.4 Some extensions
### 6.4.1 Role of institutions

Crime is related to the quality of institutions ranging from the ability of police in preventing and deterring crime and the efficacy of courts in dealing with timely verdicts of cases (Mastrobuoni, 2020). Given the high spatial variation across various states in India in the quality of courts and police, we use related measures for examining the role of institutions on the impact of road infrastructure on crime. We use pendency (share of pending to total cases in the high court), conviction (share of convicted to convicted and acquitted individuals in the high court) and prison (total number of prisons per million population of the state). We divide states into low and high categories using the group average and re-estimate the models for the two subsamples. Results have been compiled in Figure 1[13].

*<Figure 1>*

Figure 1 indicates that the negative impact of pucca roads on criminal activities is limited to states that have higher pendency, lower conviction and lower prison rates. In contrast, the effects are insignificant in the other two subsamples. Figure 1 presents an interesting pattern of how states with low quality institutions may invest resources on road infrastructure that leads to villages being better connected. In summary, road infrastructure acts as a tool for states lagging behind in institutions to catch up with states with better quality of institutions to reduce crime.

### 6.4.2. Role of inequality and employment programs

Inequality and income uncertainty are key variables that affect both infrastructure provisioning and crime. We divide our sample into two categories based on the average level of inequality at the state level. Additionally, the Government of India provides an employment guarantee scheme, named Mahatma Gandhi National Rural Employment Guarantee Scheme (MGNREGS), to tackle the uncertainty of income in rural areas by providing 100 days of employment per year to job seekers at a fixed daily wage rate. We use a similar approach of splitting our sample into two groups according to the average employment generated under MGNREGS in respective states. We present our results in Figure 2.

*<Figure 2>*

---

[13] We use crime index 1

Figure 2 shows that the negative impact of pucca roads on crime is limited to households dwelling in more unequal states and in the group of states that provide higher MGNREGS employment. These results confirm that road infrastructure is critical in combating crimes in highly unequal regions. Further, since the majority of the work under MGNREGS is related to building roads, it involves participation by local residents by getting employed in these projects. With employment opportunities directly related to building the road network, local residents face lower crimes. These extensions highlight the critical role of a well-connected road system in the rural areas of India.

## 7. Conclusion

We attempt to examine the effect of building road infrastructure on crime in rural India. We use data from the two rounds of IHDS, a nationally representative household-level survey, conducted in 2004-05 and 2011-12. Using instrumental variable estimation models, we estimate that road infrastructure has a negative impact on crime in rural India. We tease out this result by examining the underlying channels that drive it. We find that since building road infrastructure is directly related to increased provisioning of street lights and the presence of public bus services, it has an immediate impact on crime deterrence. We also find that with increased access and lower transaction costs, there are economic benefits such as better employment opportunities, increased asset holding and lower inequalities. These improve the quality of life and hence a shift away from participation in illegal activities. These effects are more pronounced if states are at a disadvantageous position in terms of institutional quality and inequality among masses.

Our results highlight the importance of building a solid infrastructure base in developing economies. In addition to improving employment opportunities, road infrastructure generates positive spillover in the form of reduced crime. This has significant policy implications in designing and implementing policies that focus on investing resources in these projects despite long gestation periods. There are few limitations of the current study. In the absence of village identifiers in the IHDS survey, PMGSY implementation and crime mapping remains unexplored. Additionally, the current paper has focused on the effect of road infrastructure on criminal activities in the rural sector. The impact of road infrastructure on the incidence of crime in urban areas is open for further research.

Despite these limitations, the current paper underscores an essential issue from a policy perspective. The ongoing COVID-19 pandemic has deepened the existing inequality of income, gender and caste in India (Deaton, 2021; Deshpande, 2021; World Bank, 2020). In such situations, the role of the government in scaling up investment in infrastructure projects like road becomes pivotal to ensure that the pandemic-induced rise in inequality is short-lived. In the rural sector, where infrastructure quality continues to be poor, building all-weather pucca roads can carry the twin benefit of improved economic opportunities due to easy access and reduced criminal activities.

**References**

Aggarwal, S. (2018). Do rural roads create pathways out of poverty? Evidence from India. *Journal of Development Economics*, *133*, 375-395.

Aggarwal, S. (2021). The long road to health: Healthcare utilization impacts of a road pavement policy in rural India. *Journal of Development Economics*, *151*, 102667.

Adukia, A., Asher, S., & Novosad, P. (2020). Educational investment responses to economic opportunity: evidence from Indian road construction. *American Economic Journal: Applied Economics*, *12*(1), 348-76.

Akee, R. K. (2006). *The Babeldaob road: the impact of road construction on rural labor force outcomes in the Republic of Palau* (No. 2452). IZA Discussion Papers.

Ansari, S., Verma, A., & Dadkhah, K. M. (2015). Crime rates in India: A trend analysis. *International criminal justice review*, *25*(4), 318-336.

Anser, M. K., Yousaf, Z., Nassani, A. A., Alotaibi, S. M., Kabbani, A., & Zaman, K. (2020). Dynamic linkages between poverty, inequality, crime, and social expenditures in a panel of 16 countries: two-step GMM estimates. *Journal of Economic Structures*, *9*, 1-25.

Asher, S., & Novosad, P. (2016). Market access and structural transformation: Evidence from rural roads in India. *Manuscript: Department of Economics, University of Oxford*.

Banerjee, A., & Somanathan, R. (2007). The political economy of public goods: Some evidence from India. *Journal of development Economics*, *82*(2), 287-314.

Becker, G. S. (1968). Crime and punishment: An economic approach. In *The economic dimensions of crime* (pp. 13-68). Palgrave Macmillan, London.

Bernburg, J. G., Thorlindsson, T., & Sigfusdottir, I. D. (2009). Relative deprivation and adolescent outcomes in Iceland: A multilevel test. *Social forces*, *87*(3), 1223-1250.

Biswas, S., & Das, U. (2021). Adding fuel to human capital: Exploring the educational effects of cooking fuel choice from rural India. *arXiv preprint arXiv:2106.01815*.

Blau, Judith R., and Peter M. Blau, "The Cost of Inequality: Metropolitan Structure and Violent Crime." American Sociological Review 47 (1982), 114-129

Block, Michael, and John Heineke, "A Labour Theoretical Analysis of Criminal Choice." American Economic Review 65 (1975), 314–325.

Bourguignon, F. (2000, December). Crime, violence and inequitable development. In Annual World Bank Conference on Development Economics 1999 (pp. 199-220). Washington, DC: World Bank.

Bourguignon, F. (2001). Crime as a social cost of poverty and inequality: a review focusing on developing countries. World Bank Discussion Papers, 171-192.

Bourguignon, F., Nuñez, J., & Sanchez, F. (2003). A structural model of crime and inequality in Colombia. Journal of the European Economic Association, 1(2-3), 440-449.

Brown, R., Montalva, V., Thomas, D., & Velásquez, A. (2019). Impact of violent crime on risk aversion: Evidence from the Mexican drug war. *Review of Economics and Statistics*, *101*(5), 892-904.

Buonanno, P., & Montolio, D. (2008). Identifying the socio-economic and demographic determinants of crime across Spanish provinces. *International Review of law and Economics*, *28*(2), 89-97.

Cahill, M. E., & Mulligan, G. F. (2003). The determinants of crime in Tucson, Arizona1. *Urban Geography*, *24*(7), 582-610.

Chiu, W. Henry, and Paul Madden, "Burglary and Income Inequality," Journal of Public Economics 69 (1998), 123–141.

Coffin, A. W. (2007). From roadkill to road ecology: a review of the ecological effects of roads. *Journal of transport Geography*, *15*(5), 396-406.

Conley, T. G., Hansen, C. B., & Rossi, P. E. (2012). Plausibly exogenous. *Review of Economics and Statistics*, *94*(1), 260-272.

Cook, P. J., Machin, S., Marie, O., & Mastrobuoni, G. (2013). Crime economics in its fifth decade. Lessons from the economics of crime: What reduce offending, 1-16.

Davies, T., & Johnson, S. D. (2015). Examining the relationship between road structure and burglary risk via quantitative network analysis. *Journal of Quantitative Criminology*, *31*(3), 481-507.

Das, U. (2018). Rural employment guarantee programme in India and its impact on household educational decisions: A focus on private coaching. *Journal of International Development*.

Deaton, A. (2021). Covid-19 and global income inequality (No. w28392). National Bureau of Economic Research.

Deshpande, A. (2021). How India's Caste Inequality Has Persisted—and Deepened in the Pandemic. Current History, 120(825), 127-132.

Dutta, M., & Husain, Z. (2009). Determinants of crime rates: Crime Deterrence and Growth in post-liberalized India.

Ehrlich, Isaac, "Participation in Illegitimate Activities: A Theoretical and Empirical Investigation." Journal of Political Economy 81 (1973), 521–565.

Fajnzylber, P., Lederman, D., & Loayza, N. (1998). *Determinants of crime rates in Latin America and the world: an empirical assessment*. World Bank Publications.

Fajnzylber, P., Lederman, D., & Loayza, N. (2002). Inequality and violent crime. *The journal of Law and Economics*, *45*(1), 1-39.

Ferreira, F. H. (1995). Roads to equality: wealth distribution dynamics with public-private capital complementarity. *LSE STICERD Research Paper No. TE286*.

Forman, R. T., & Alexander, L. E. (1998). Roads and their major ecological effects. *Annual review of ecology and systematics*, *29*(1), 207-231.

Hughes, D. W. (1996). The Effects of Infrastructure Development on Crime in Rural Areas: A Case Study of the A55 Coastal Expressway in North Wales. *Cambrian L. Rev.*, *27*, 33.

İmrohoroĝlu, Ayse, Antonio Merlo, and Peter Rupert. "Understanding the determinants of crime." *Journal of Economics and Finance* 30.2 (2006): 270-284.

Iyer, L., Mani, A., Mishra, P., & Topalova, P. (2012). The power of political voice: women's political representation and crime in India. *American Economic Journal: Applied Economics*, *4*(4), 165-93.

Jha, B. (2006). Rural non-farm employment in India: Macro-trends, micro-evidences and policy options. *Delhi: Agricultural Economics Unit, Institute of Economic Growth*.

Kang, S. (2016). Inequality and crime revisited: Effects of local inequality and economic segregation on crime. *Journal of Population Economics*, *29*(2), 593-626.

Khanani, R. S., Adugbila, E. J., Martinez, J. A., & Pfeffer, K. (2021). The impact of road infrastructure development projects on local communities in peri-urban areas: the case of Kisumu, Kenya and Accra, Ghana. *International Journal of Community Well-Being*, *4*(1), 33-53.

Khandker, S. R., & Mundial, B. (1989). *Improving rural wages in India* (No. 276). Population and Human Resources Department, World Bank.

Kelly, M. (2000). Inequality and crime. *Review of economics and Statistics*, *82*(4), 530-539.

Blau, Judith R., and Peter M. Blau, "The Cost of Inequality: Metropolitan Structure and Violent Crime." American Sociological Review 47 (1982), 114-129

Lei, L., Desai, S., & Vanneman, R. (2019). The impact of transportation infrastructure on women's employment in India. *Feminist economics*, *25*(4), 94-125.

Mahajan, K., & Sekhri, S. (2020). *Access to toilets and violence against women* (No. 44).

Mastrobuoni, G. (2020). Crime is terribly revealing: Information technology and police productivity. The Review of Economic Studies, 87(6), 2727-2753.

Merton, Robert, "Social Structure and Anomie," American Sociological Review 3 (1938), 672-682

O'Mahony, Margaret. "The Relative Influence of Proximity to Fast Road Infrastructure, Accessibility, and Deprivation on Crime." *Journal of Advanced Transportation* 2018 (2018).

Paul, M. J., & Meyer, J. L. (2001). Streams in the urban landscape. *Annual review of Ecology and Systematics*, *32*(1), 333-365.

Perkins, C., Steinbach, R., Tompson, L., Green, J., Johnson, S., Grundy, C., ... & Edwards, P. (2015). The effect of reduced street lighting on crime and road traffic injuries at night in England and Wales: a controlled interrupted time series analysis. In *What is the effect of reduced street lighting on crime and road traffic injuries at night? A mixed-methods study*. NIHR Journals Library.

Prasad, K. (2013). A comparison of victim-reported and police-recorded crime in India. *Economic and Political Weekly*, 47-53.

Samanta, P. K. (2015). Development of rural road infrastructure in India. *Pacific business review international*, *7*(11), 86-93.

Sawada, Y., Shoji, M., Sugawara, S., & Shinkai, N. (2014). The role of infrastructure in mitigating poverty dynamics: The case of an irrigation project in Sri Lanka. *The BE Journal of Economic Analysis & Policy*, *14*(3), 1117-1144.

Slabbekoorn, H., & Peet, M. (2003). Birds sing at a higher pitch in urban noise. *Nature*, *424*(6946), 267-267.

Van der Hoeven, R. (2019). Income Inequality in Developing Countries, Past and Present. In The Palgrave Handbook of Development Economics (pp. 335-376). Palgrave Macmillan, Cham.

**List of Figures for the manuscript**

**Figure 1- Influence of institutional factors on the main effect**

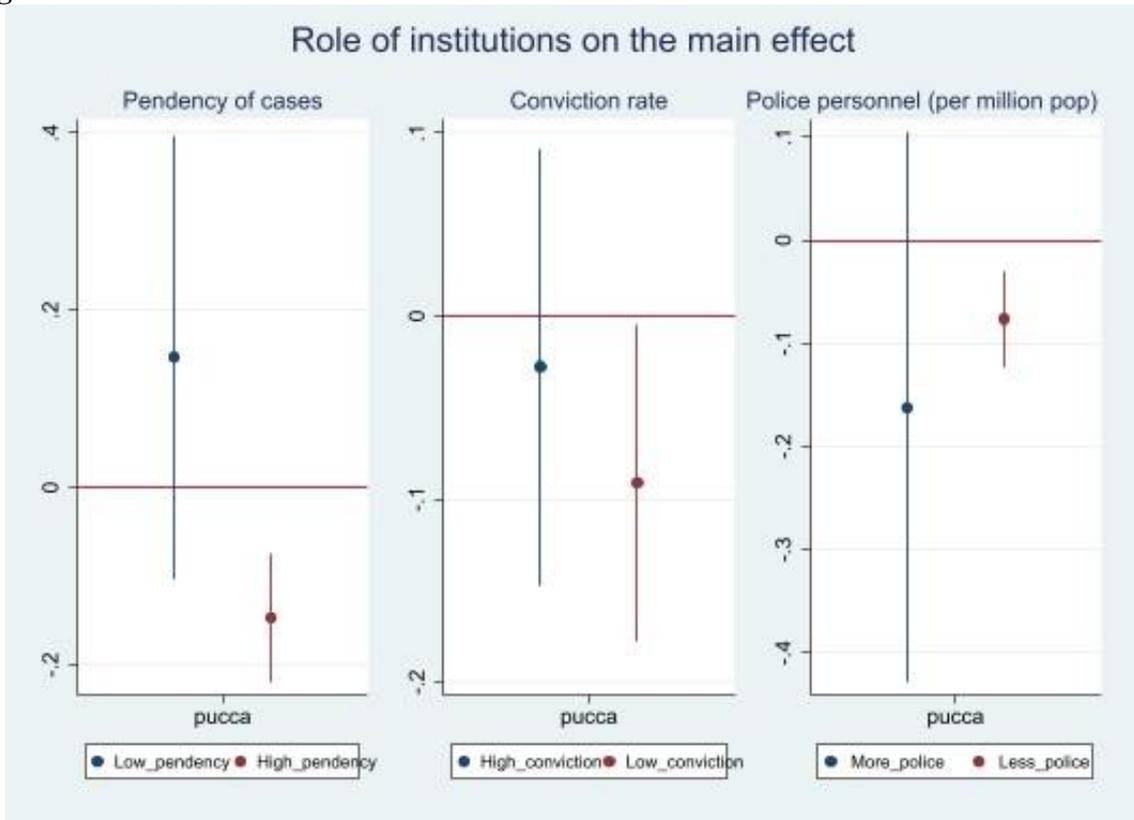

**Figure 2- Influence of casual employment and inequality on the main effect**

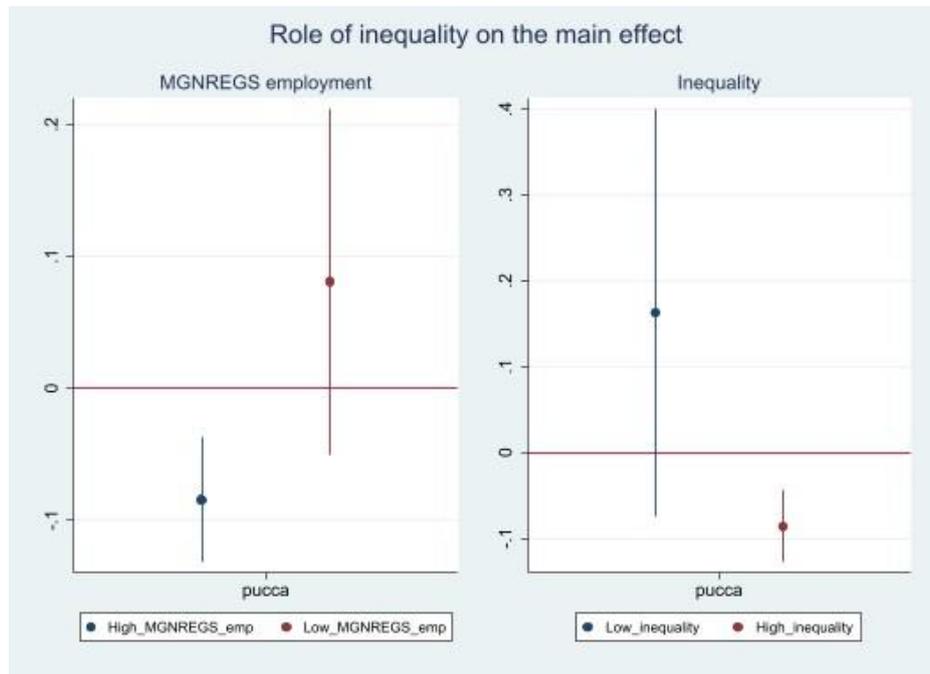

**List of Tables**

**Table 1- Summary statistics of variables**

| Variable | Measure | Obs | Mean | Std. Dev. | Min | Max |
|---|---|---|---|---|---|---|
| *Crime incidence* | | | | | | |
| Theft | Dummy variable for anything stolen in the last 12 months | 54449 | 0.03 | 0.19 | 0 | 1 |
| Attack | Dummy variable for getting attacked or threatened in the last 12 months | 54449 | 0.02 | 0.15 | 0 | 1 |
| Harassment | Dummy variable for unmarried women getting harassed in the last 12 months | 54525 | 0.15 | 0.35 | 0 | 1 |
| Breaking into home | Dummy variable for someone breaking into home in the last 12 months | 54449 | 0.02 | 0.09 | 0 | 1 |
| Crime index 1 | Simple average of all the four criminal activities mentioned above | 54436 | 0.03 | 0.10 | 0 | 1 |
| Crime index 2 | Dummy variable that takes a unit value if any crime is committed | 54525 | 0.18 | 0.39 | 0 | 1 |
| *Village level infrastructure and composition measures* | | | | | | |
| *Direct road measures* | | | | | | |
| Pucca road | Dummy variable for an all-weather pucca road in the village | 54525 | 0.75 | 0.43 | 0 | 1 |
| *Other infrastructure measures* | | | | | | |
| Presence of police station | Ln(1+Distance from the police station) | 54505 | 2.25 | 0.62 | 0 | 4.15 |
| Street light program | Dummy variable for villages that have a public program that spends on it | 54491 | 0.38 | 0.51 | 0 | 1 |
| Bus stop | Ln(1+ distance of a bus stop from the village) | 54525 | 1.14 | 0.61 | 0 | 3.73 |
| *Village composition measures* | | | | | | |
| Caste based land ownership diff | Difference in share of land ownership between general caste and others | 54525 | 0.10 | 0.38 | 0 | 1 |
| Religion based land ownership diff | Difference in share of land ownership between majority religion and others | 54525 | 0.51 | 0.48 | 0 | 1 |
| Wage diff between two seasons | (Wage difference between the kharif and rabi season for men)/average wage | 52380 | 1.55 | 0.59 | -0.08 | 6 |
| Migrants from outside the village | How many people came to work during the last year? (1- less than 20, 2- more than 20) | 54525 | 0.80 | 0.94 | 0 | 2 |
| *Confidence in institutions* | | | | | | |
| Police | Dummy variable for confidence in courts | 54192 | 0.70 | 0.43 | 0 | 1 |
| Panchayat | Dummy variable for confidence in panchayat | 54029 | 0.80 | 0.39 | 0 | 1 |
| Courts | Dummy variable for confidence in courts | 52955 | 0.80 | 0.30 | 0 | 1 |
| *Household level factors* | | | | | | |
| Household income | Share of the assets owned by the hhd to the tot number of assets mentioned | 54525 | 0.31 | 0.14 | 0 | 1 |
| Employment ratio | Share of adults in the household that are work > 240 hrs in a year | 53496 | 0.67 | 0.30 | 0 | 1 |
| Household size | Log (total number of individuals in the household) | 54525 | 5.18 | 2.55 | 1 | 38 |

**Table 2- Mean difference according to road infrastructure in the village**

| Variables | Pucca roads | No pucca roads | Difference |
|---|---|---|---|
| *Types of criminal activity* | | | |
| Crime index 1 | 0.023 | 0.028 | -0.005*** |
| Crime index 2 | 0.188 | 0.188 | 0.000 |
| Theft | 0.036 | 0.045 | -0.008*** |
| Attack | 0.023 | 0.030 | -0.007*** |
| Harassment | 0.145 | 0.150 | -0.004* |
| Breaking into home | 0.009 | 0.010 | 0.001 |
| *Other facilities and economic outcomes* | | | |
| Street lights | 0.437 | 0.236 | 0.201*** |
| Bus stop distance | 1.043 | 1.471 | -0.428*** |
| Employment ratio | 0.759 | 0.684 | 0.074*** |
| Household income | 0.327 | 0.279 | 0.048*** |

**Table 3- Effects of road infrastructure on crime in India**

| Dependent Variable- Crime index | Model 1 | Model 2 | Model 3 | Model 4 | Model 5 | Model 6 |
|---|---|---|---|---|---|---|
| *Road infrastructure* | | | | | | |
| Pucca road | -0.009*** | -0.076*** | -0.027*** | -0.047** | -0.123* | -0.341** |
|  | (0.002) | (0.013) | (0.005) | (0.024) | (0.065) | (0.203) |
| *Confidence in institutions* | | | | | | |
| Confidence in police | -0.009*** | -0.002 | -0.025*** | -0.009*** | -0.009*** | -0.096*** |
|  | (0.001) | (0.004) | (0.004) | (0.001) | (0.002) | (0.020) |
| Confidence in panchayat | -0.009*** | 0.007 | -0.025*** | -0.008*** | -0.008*** | -0.084*** |
|  | (0.002) | (0.004) | (0.005) | (0.002) | (0.002) | (0.024) |
| Confidence in courts | -0.006*** | -0.005 | 0.014** | -0.007*** | -0.006*** | 0.038 |
|  | (0.002) | (0.005) | (0.006) | (0.002) | (0.002) | (0.027) |
| *Village characteristics* | | | | | | |
| Presence of police station | -0.000*** | -0.004*** | -0.001*** | -0.001*** | -0.001*** | -0.008*** |
|  | (0.000) | (0.000) | (0.000) | (0.000) | (0.000) | (0.002) |
| Caste based land ownership diff |  | 0.003 | 0.030*** |  | 0.010*** | 0.110*** |
|  |  | (0.008) | (0.009) |  | (0.003) | (0.037) |
| Religion based land ownership diff |  | 0.023*** | 0.028*** |  | 0.011*** | 0.137*** |
|  |  | (0.008) | (0.008) |  | (0.003) | (0.032) |
| Wage diff between two seasons |  | 0.019*** | 0.010*** |  | 0.004*** | 0.062*** |
|  |  | (0.003) | (0.003) |  | (0.002) | (0.015) |
| Migrants from outside the village |  | 0.032*** | 0.007*** |  | 0.006*** | 0.070*** |
|  |  | (0.002) | (0.002) |  | (0.002) | (0.020) |
| *Household level characteristics* | | | | | | |
| Income | -0.000 | 0.167*** | -0.011 | 0.007 | 0.014 | 0.175 |
|  | (0.005) | (0.013) | (0.015) | (0.007) | (0.012) | (0.132) |
| Household size | 0.000 | -0.001 | -0.002* | 0.000 | -0.001* | -0.010** |
|  | (0.000) | (0.001) | (0.001) | (0.000) | (0.000) | (0.005) |
| Elderly |  | 0.010** | 0.022*** |  | 0.005*** | 0.072*** |
|  |  | (0.005) | (0.005) |  | (0.002) | (0.028) |
| *Other controls* | | | | | | |
| Religion FE | Yes | Yes | Yes | Yes | Yes | Yes |
| Caste FE | Yes | Yes | Yes | Yes | Yes | Yes |
| Year FE | Yes | Yes | Yes | Yes | Yes | Yes |
| District FE | Yes | Yes | Yes | Yes | Yes | Yes |
| Constant | 0.085*** | 0.200*** | 0.263*** | 0.103*** | 0.104*** | -0.398* |
|  | (0.011) | (0.036) | (0.038) | (0.016) | (0.018) | (0.241) |
| Observations | 50727 | 48605 | 48605 | 50727 | 48605 | 48505 |
| R-squared | 0.10 | 0.30 | 0.11 | 0.10 | 0.08 |  |

Note: Models 1, 2 and 3 are the results from linear regression models where as Models 4, 5 and 6 from the second stage of the instrumental variable regression models. We use *ivreg2* for Models 4 and 5 and *ivprobit* for Model 6. The values indicate the coefficients of each of the covariates. Robust standard errors are presented in parentheses. *** p<0.01, ** p<0.05, * p<0.1.

**Table 4- Estimates of the instruments obtained from the first stage regressions**

|                                          | Model 1   | Model 2   | Model 3   |
|------------------------------------------|-----------|-----------|-----------|
| *Instruments*                            |           |           |           |
| Proportion of households with piped water| 0.059***  | 0.023***  | 0.023***  |
|                                          | (0.004)   | (0.004)   | (0.004)   |
| Other exogenous variables                | Yes       | Yes       | Yes       |
| Religion FE                              | Yes       | Yes       | Yes       |
| Caste FE                                 | Yes       | Yes       | Yes       |
| Year FE                                  | Yes       | Yes       | Yes       |
| District FE                              | Yes       | Yes       | Yes       |
| No. of observations                      | 50727     | 48605     | 48505     |
| *Tests of instrument validity*           |           |           |           |
| *F test of excluded instrument*          |           |           |           |
| Sanderson Windmeijer test                | 184.69    | 28.80     | 28.90     |
|                                          | (0.000)   | (0.000)   | (0.000)   |
| *Under-identification test*              |           |           |           |
| Anderson Canon. Corr. LM statistic       | 4.59      | 4.59      | 4.33      |
|                                          | (0.032)   | (0.032)   | (0.037)   |
| *Weak- identification test*              |           |           |           |
| Cragg-Donald wald statistic              | 179.34    | 27.80     | 27.89     |
| 10% maximum IV value                     | 16.38     | 16.38     | 16.38     |

Note: The top panel of this table presents estimates of the instrument from the first stage of Models 4, 5 and 6 from Table 1. Standard errors are reported within parentheses. *, ** and *** indicate significance at 10, 5 and 1 percent, respectively. The bottom panel presents the tests of instrument validity and relevance. The values indicate the test statistic and p-values are reported in parentheses.

**Table 5- Effect of road infrastructure on various types of criminal activity**

|  | Theft | Attack | Harassment | Breaking into home |
|---|---|---|---|---|
| *Road infrastructure* | | | | |
| Pucca road | -0.822** | 0.301 | -0.642** | 0.202* |
|  | (0.432) | (0.592) | (0.344) | (0.124) |
| Other controls | Yes | Yes | Yes | Yes |
| Religion FE | Yes | Yes | Yes | Yes |
| Caste FE | Yes | Yes | Yes | Yes |
| Year FE | Yes | Yes | Yes | Yes |
| District FE | Yes | Yes | Yes | Yes |
| Observations | 46284 | 40672 | 50524 | 34290 |

Note: This table presents the regression results for determinants of various types of criminal activities. Each of the models use *ivprobit* since the dependent variables are dichotomous. The values indicate the coefficients of each of the covariates. Robust standard errors are presented in parentheses. *** $p<0.01$, ** $p<0.05$, * $p<0.1$

**Table 6- Bounds from the plausibly exogenous estimations**

| Dependent variable | Crime 1 | Crime 2 |
|---|---|---|
| $\hat{\gamma}$ | -0.002** | -0.009** |
|  | (0.001) | (0.004) |
| All controls | Yes | Yes |
| Fixed effects | Yes | Yes |
| Observations | 48605 | 48605 |
| *γ values (95% of estimated values)* | | |
| Minimum γ value | -0.002 | -0.009 |
| Maximum γ value | 0 | 0 |
| *Pucca road* | | |
| Lower bound for Pucca road | -0.831 | -2.392 |
| Upper bound for Pucca road | -0.345 | -0.884 |

Note: All covariates and dummy variables are included in both the models. *Plausexog* command has been used in STATA 16 with the Union of Confidence Interval (UCI) approach.

**Table 7- Potential channels of how road infrastructure reduces criminal activities**

|  | Direct effect | | | Indirect effect | | |
|---|---|---|---|---|---|---|
|  | Street light | Bus stop presence | Employment status | Household income | Land ownership diff- caste | Land ownership diff- religion |
| *Road infrastructure* | | | | | | |
| Pucca road | 1.780*** | 1.727*** | 0.308* | 0.043** | -1.178*** | -1.021*** |
|  | (0.155) | (0.131) | (0.164) | (0.022) | (0.258) | (0.237) |
| Other controls | Yes | Yes | Yes | Yes | Yes | Yes |
| Religion FE | Yes | Yes | Yes | Yes | Yes | Yes |
| Caste FE | Yes | Yes | Yes | Yes | Yes | Yes |
| Year FE | Yes | Yes | Yes | Yes | Yes | Yes |
| District FE | Yes | Yes | Yes | Yes | Yes | Yes |
| Observations | 48658 | 48691 | 47767 | 48,661 | 48,658 | 48,658 |
| R-squared | 0.70 | 0.49 | 0.10 | 0.11 | 0.17 | 0.60 |

Note: This table presents the regression results for the effect of road infrastructure on presence of street lights, bus stop within the village, employment status, household income and difference in land ownership in the village at the caste and religion level. We use *ivreg2* to obtain the results. The values indicate the coefficients of each of the covariates. Robust standard errors are presented in parentheses. *** $p<0.01$, ** $p<0.05$, * $p<0.1$